\documentclass[fleqn,usenatbib]{mnras}

\usepackage{newtxtext,newtxmath}

\usepackage[T1]{fontenc}

\DeclareRobustCommand{\VAN}[3]{#2}
\let\VANthebibliography\thebibliography
\def\thebibliography{\DeclareRobustCommand{\VAN}[3]{##3}\VANthebibliography}

\usepackage{graphicx}	

\usepackage{color}

\title[Feeding the beast: Disrupting asteroids]{Formation of eccentric gas discs from sublimating or partially disrupted asteroids orbiting white dwarfs}

\author[D. Trevascus et al.]{
David Trevascus$^{1}$\thanks{E-mail: dtre10@student.monash.edu (KTS)},
Daniel J. Price$^{1}$,
Rebecca Nealon$^{2,3}$, 
David Liptai$^{1}$,
Christopher J. Manser$^{4}$ and Dimitri Veras$^{2,3}$
\\
$^{1}$School of Physics and Astronomy, Monash University, VIC 3800, Australia \\
$^{2}$Department of Physics, University of Warwick, Coventry CV4 7AL, UK \\
$^{3}$Centre for Exoplanets and Habitability, University of Warwick, Coventry CV4 7AL, UK \\
$^{4}$Astrophysics Group, Department of Physics, Imperial College London, Prince Consort Rd, London, SW7 2AZ, UK
}

\date{Accepted XXX. Received YYY; in original form ZZZ}

\pubyear{2021}

\defcitealias{Manser_2019}{M19}

\begin{document}
\label{firstpage}
\pagerange{\pageref{firstpage}--\pageref{lastpage}}
\maketitle

\begin{abstract}
Of the 21 known gaseous debris discs around white dwarfs, a large fraction of them display observational features that are well described by an eccentric distribution of gas. In the absence of embedded objects or additional forces, these discs should not remain eccentric for long timescales, and should instead circularise due to viscous spreading. The metal pollution and infrared excess we observe from these stars is consistent with the presence of tidally disrupted sub-stellar bodies. We demonstrate, using smoothed particle hydrodynamics, that a sublimating or partially disrupting planet on an eccentric orbit around a white dwarf will form and maintain a gas disc with an eccentricity within 0.1 of, and lower than, that of the orbiting body. We also demonstrate that the eccentric gas disc observed around the white dwarf SDSS J1228+1040 can be explained by the same hypothesis.
\end{abstract}




\section{Introduction}

Heavy metal pollution is observed in 25-50 percent of white dwarfs \citep{Zuckerman_2003, Zuckerman_2010, Koester_2014}. The diffusion timescale of these elements from the atmosphere of these stars is short in comparison to their age \citep{Fontaine_1979} so material must be fed to these stars to produce the lines. Remnant planetary systems are the likely origin \citep[e.g.][]{Debes_2002, Bonsor_2011, Frewen_2014, Smallwood_2018}. Bodies from these systems tidally disrupt and form dusty debris discs \citep{Veras_2014, Malamud_2020}.

In around 21 of the white dwarfs which are host to dusty debris discs we have also observed a gaseous component to the disc \citep{Gansicke_2006, Gansicke_2007, Gansicke_2008, Gansicke_2011, Farihi_2012, Melis_2012, Wilson_2014, Dennihy_2020, GentileFusillo_2020, Melis_2020}. The presence of these discs is inferred through the detection of the double-peaked Ca II emission triplet ($\sim8600$ \AA) in the spectrum of the system \citep{Horne_1986}. We have observed continuous changes in shape of the emission features, over timescales of years, in at least six of these systems \citep{Wilson_2015, Manser_2016a, Manser_2016b, Dennihy_2018, Dennihy_2020}, and of the absorption features in at least one other system \citep{Cauley_2018}, suggesting a stable eccentric disc structure.

Of particular note is the gas disc orbiting the white dwarf SDSS J122859.93+104032.9 (from here SDSS J1228+1040). \cite{Gansicke_2006} fit the evolution of the line profiles to a disc with an eccentricity of 0.021. Optical spectroscopy reported by \cite{Manser_2016a} revealed a continuous variation in the double-peaked Ca II emission lines from redshift to blueshift over a period of 12 years. The authors speculated that the slow shifting of the spectral lines was due to material precessing around the white dwarf, with a timescale consistent with apsidal advance around the white dwarf due to General Relativity. By fitting a precession period of approximately 27 years, they produced a tomogram map \citep{Marsh_1988} showing the structure of the gas disc in velocity space (reproduced in our Figure~\ref{fig:tom_comp}). \citet{Cauley_2018} and \citet{Dennihy_2018} similarly used general relativistic precession of an eccentric gas disc  to explain the variations in spectral lines from a number of other white dwarfs, a process described analytically by \cite{Miranda_2018}.

Even more intriguing, subsequent detection of short-timescale periodic rocking, over a period of approximately 2 hrs, in the spectral lines of SDSS J1228+1040 by \cite{Manser_2019} (hereafter~\citetalias{Manser_2019}) hinted at a non-disintegrating orbiting planetesimal. Other planetesimals detected orbiting close to white dwarfs are thought to be disrupting \citep{Vanderburg_2015, Vanderbosch_2020, Guidry_2020}, suggesting that this body must have increased internal strength or higher density to prevent complete disruption.

Several studies examined the formation of dust discs from the tidal disruption of solid bodies around white dwarfs \citep{Veras_2014, Veras_2015, Kenyon_2017a, Kenyon_2017b,  Malamud_2020}. \cite{Rafikov_2011} and \cite{Metzger_2012} studied gas disc formation through dust sublimation, while \cite{Kenyon_2017a, Kenyon_2017b} assumed a collisional cascade. But none of these studies explain the observed eccentricity in the discs. A key problem is that gas would be expected to circularise due to viscous spreading \citep{Lynden-Bell_1974} rather than remain eccentric (although the circularisation timescale may be long, see \citealt{Nixon_2020}).

\begin{figure*}
	\includegraphics[width=\textwidth]{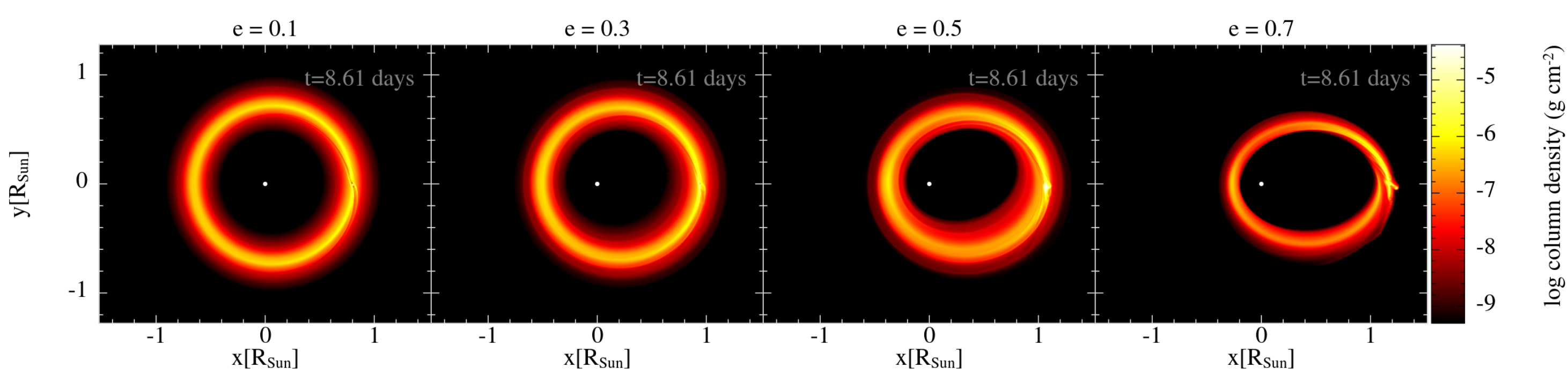}
    \caption{Column density of material in simulated gas discs formed by bodies with orbital $e=0.1, 0.3, 0.5$, and $0.7$. The discs are shown at a single point in time after 100 orbits, showing the disc eccentricity is sustained over at least this length of time. The white dot at the origin in each image shows the position of the white dwarf relative to the disc. 
    }
    \label{fig:splash_comp}
\end{figure*}

Between 4\% and 30\% percent of white dwarfs with dust discs show the presence of circumstellar gas \citep{Dennihy_2020, GentileFusillo_2020, Manser_2020, Melis_2020}.


In this Letter, we hypothesise a sublimating or partially disrupted body as the source of the eccentric gas disc, motivated by the observations of SDSS J1228+1040 described above \citep{Manser_2016a, Manser_2019, Manser_2020}. We show that such a body could drive disc eccentricity, while continuously supplying the disc with material. We test this hypothesis by simulating the emission of gas from a point mass on an eccentric orbit around the white dwarf to show the formation and stability of an eccentric gas disc. 


\begin{figure*}
	\includegraphics[width=\textwidth]{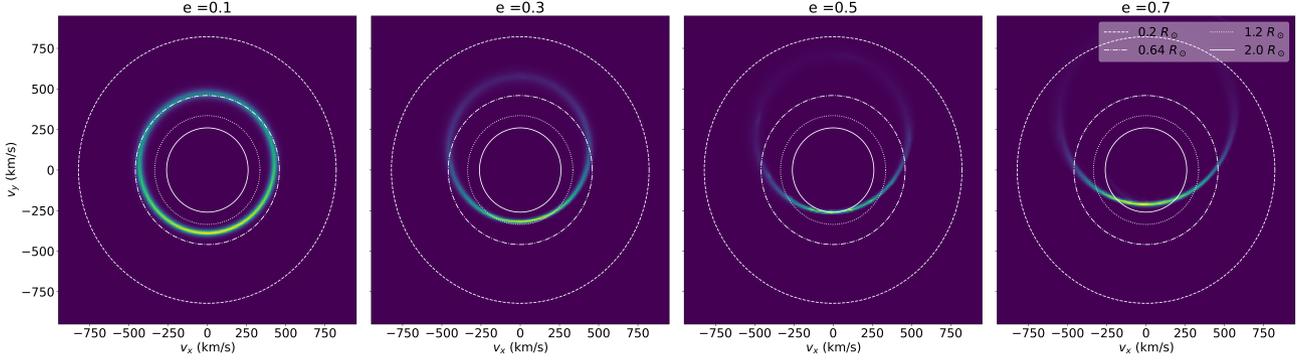}
    \caption{Synthetic tomograms corresponding to simulations shown in Figure \ref{fig:splash_comp}. We show a time average over the 100th orbit of the point mass. The circular white lines on each plot indicate circular Keplerian orbits with semi-major axes of 0.2 R$_\odot$, 0.64 R$_\odot$, 1.2 R$_\odot$ and 2.0 R$_\odot$ (taken from the outermost circle and moving inwards), assuming a 90$^\circ$ inclination. Regions of highest density are shown in yellow, while low density regions are light blue.}
    \label{fig:tom_ecc_comp}
\end{figure*}

\section{Methods}
\label{sec:methods}
We modelled the formation and evolution of a gas disc around a white dwarf using the smoothed particle hydrodynamics \citep[SPH; e.g.][]{Lucy_1977,Gingold_1977} code {\sc phantom} \citep{Price_2018}. We adopted the modified Newtonian potential from \citet{Tejeda_2013}, implemented in {\sc phantom} by \citet{Bonnerot_2016} to model the general relativistic potential around a white dwarf. We set a point mass (sink particle) on an eccentric orbit in this potential. For our initial set of calculations, we set the initial semi-major axis, $a_0 = 0.73$\,R$_\odot$, corresponding to the 123.4 minute periodic signal from \citetalias{Manser_2019}, and eccentricities $e=0.1,0.3,0.5$ and $0.7$, respectively.

We set the mass of the orbiting body to 0.1 Ceres masses, the maximum mass of the disintegrating body orbiting the white dwarf WD 1145+017 as determined by \cite{Gurri_2017}. We set the injection radius, $R_{\rm inj} = 2338.3$ km (relative to the centre of the orbiting body), which is 0.01 times the periastron radius of a body orbiting with $e=0.54$, the inferred eccentricity of the planetesimal discovered by \citetalias{Manser_2019}, and semi-major axis $a_0$.

We assumed a white dwarf of $0.705$\,M$_\odot$, as determined for SDSS J1228+1040 by \citet{Koester_2014}. For computational efficiency, we set an accretion radius for the white dwarf of $0.2$\,R$_\odot$, interior to which particles were deleted from the simulation.

We set the mass injection rate of gas $\dot{M} = 5\times 10^8$\,g\,s$^{-1}$, assuming that the generation of gas occurs at the same rate as the accretion rate of metal pollution measured for SDSS J1228+1040 by \cite{Gansicke_2012}. As the light flux received by the orbiting body increases, the amount of gas released through volatiles should also increase. Given that the light flux is proportional to $\frac{1}{r^2}$, where $r$ is the distance between the orbiting body and the white dwarf, we set the mass injection rate of gas from the orbiting body to be given by
\begin{equation} \label{eq:mass inject}
    \dot{M} = 5\times 10^8 \text{g\,s}^{-1} \left(\frac{a_0}{r}\right)^2.
\end{equation}

We employed a gas particle injection rate of $\sim 10^5$ particles per orbit, with the injected mass divided evenly between the particles, giving $10^7$ particles in the simulation at the time shown in Figure~\ref{fig:splash_comp}. That is, we injected SPH particles in a uniform random distribution from $R_{\rm inj}$ with a mass per particle set to $3.71\times 10^{12}$ g. Gravitational forces between gas particles were neglected.

We assumed an isothermal equation of state with $T=5000$K, as determined by \cite{Melis_2010} and \cite{Hartmann_2016}. Assuming a mean molecular weight of $\mu=0.6$, this results in a sound speed $c_{\rm s} = 8.3$\,km/s and hence a disc with an aspect ratio $H/R \equiv c_{\rm s}/(\Omega_0 a_0) = 0.02$, i.e. a thin disc. We model a disc viscosity by the standard procedure of using the SPH shock viscosity to mimic the \cite{Shakura_1973} prescription, adopting $\alpha_{\rm av}=1$ corresponding to $\alpha_{\rm ss} \approx 0.1 \alpha_{\rm av} \langle h\rangle/H \approx 0.05$ following the prescription in \citet{Lodato_2010}.

\subsection{Synthetic Observations and Fitting Orbits}
We computed 2D histograms, binning the SPH particles by velocities $v_x$ and $v_y$, to produce synthetic tomograms. These were time-averaged from ten snapshots spaced evenly in time over a single orbit, and convolved with a Gaussian beam of width 9.06 km~s$^{-1}$ in order to match the resolution of the observational data.

In the above we effectively assume that the gas disc around the white dwarf is optically thin. Spectral lines for most white dwarf gaseous debris discs are optically thick \citep{Gansicke_2006}, but this should not alter the effects of time-averaging. The assumption of optical thinness simplifies the equation of radiative transfer,
\begin{equation} \label{eq:rad. transfer}
    I_\nu (s) = I_\nu (s_0)e^{-\tau_\nu (s_0,s)} + \int_{s_0}^s j_\nu (s')e^{-\tau_\nu (s',s)}ds',
\end{equation}
so that the optical depth $\tau_\nu$, and therefore the column density of disc material, is proportional to the intensity of the spectral lines $I_\nu$. In this equation, $\nu$ is the frequency of the light travelling through the disc, $s$ is the distance travelled by light through the disc from a reference point $s_0$, and $j_\nu$ is the emissivity.

We fitted a test particle orbit to the gas disc in each simulation. These orbits were characterised by a semi-major axis $a$, eccentricity $e$, and an additional phase to account for rotation in the $x,y$ plane.

We transformed the velocity distribution of gas particles in each disc into polar coordinates and fit the data in each angular bin to a Gaussian to find the velocity $|\mathbf{v}|_{\rm dat, n}$ with the highest density of particles for that bin. We used the standard deviation $\sigma_{\rm dat, n}$ of the Gaussian fit as the uncertainty for this value.

The closest fitting orbit was determined by minimising the negative of a logarithmic likelihood function, given by
\begin{equation} \label{eq:log likelihood}
    \ln p\left(|\mathbf{v}|_{\rm dat} \:|\: |\mathbf{v}|,\sigma_{\rm dat}\right) = -\frac{1}{2} \sum_n \left[\frac{|\mathbf{v}|_{\rm dat, n} - |\mathbf{v}|_{\rm n}}{\sigma_{\rm dat, n}} + \ln(\sigma_{\rm dat, n})\right]
\end{equation}
where $|\mathbf{v}|_{\rm n}$ are the velocity magnitudes produced for a given set of orbital parameters. Uncertainties on the best fitting parameters were found using the Markov chain Monte Carlo (MCMC) Python algorithm {\sc emcee} \citep{Foreman_Mackey_2013}.

\section{Results}
\label{sec:results}
Figure \ref{fig:splash_comp} shows column density in our simulated gas discs generated from a body orbiting with $e=0.1, 0.3, 0.5$, and $0.7$ at 100 orbits (of the point mass) after the start of the simulations. Figure \ref{fig:tom_ecc_comp} shows the corresponding tomograms, time averaged over the 100th orbit.

Our orbital fits applied to the gas discs show that they have eccentricities of $0.097\pm0.003$, $0.288\pm0.003$, $0.468\pm0.005$, and $0.626\pm^{0.002}_{0.003}$ respectively. These eccentricities are all below that of the orbiting body, suggesting some degree of circularisation, but they are all within 0.08 of the orbiting body. This confirms our basic hypothesis that a sublimating or partially disrupting body can create and maintain an eccentric gas disc over a timescale that is long compared with the orbital period. In all four of these simulations gas passes inwards through our simulated accretion radius, providing a source for potential white dwarf pollutants. Ideally one would continue these simulations over the full precession period, but this proved prohibitive in terms of computational cost.

Our results suggest that a large undisrupted body can produce gas discs with the particular structure that we observe. Since the gas is created by the orbiting body, there is an overdensity (see Figure \ref{fig:splash_comp}) that moves with the orbital phase of the planet. However, this gas also spends more time further away from the white dwarf, following Kepler's 2nd law. We see this in the time averaged tomograms shown in Figure~\ref{fig:tom_ecc_comp}, where there is an increased concentration of gas at apastron spanning $\sim135$ degrees across all eccentricities.

\subsection{Comparison with SDSS J1228+1040}
\begin{figure*}
	\includegraphics[width=\textwidth]{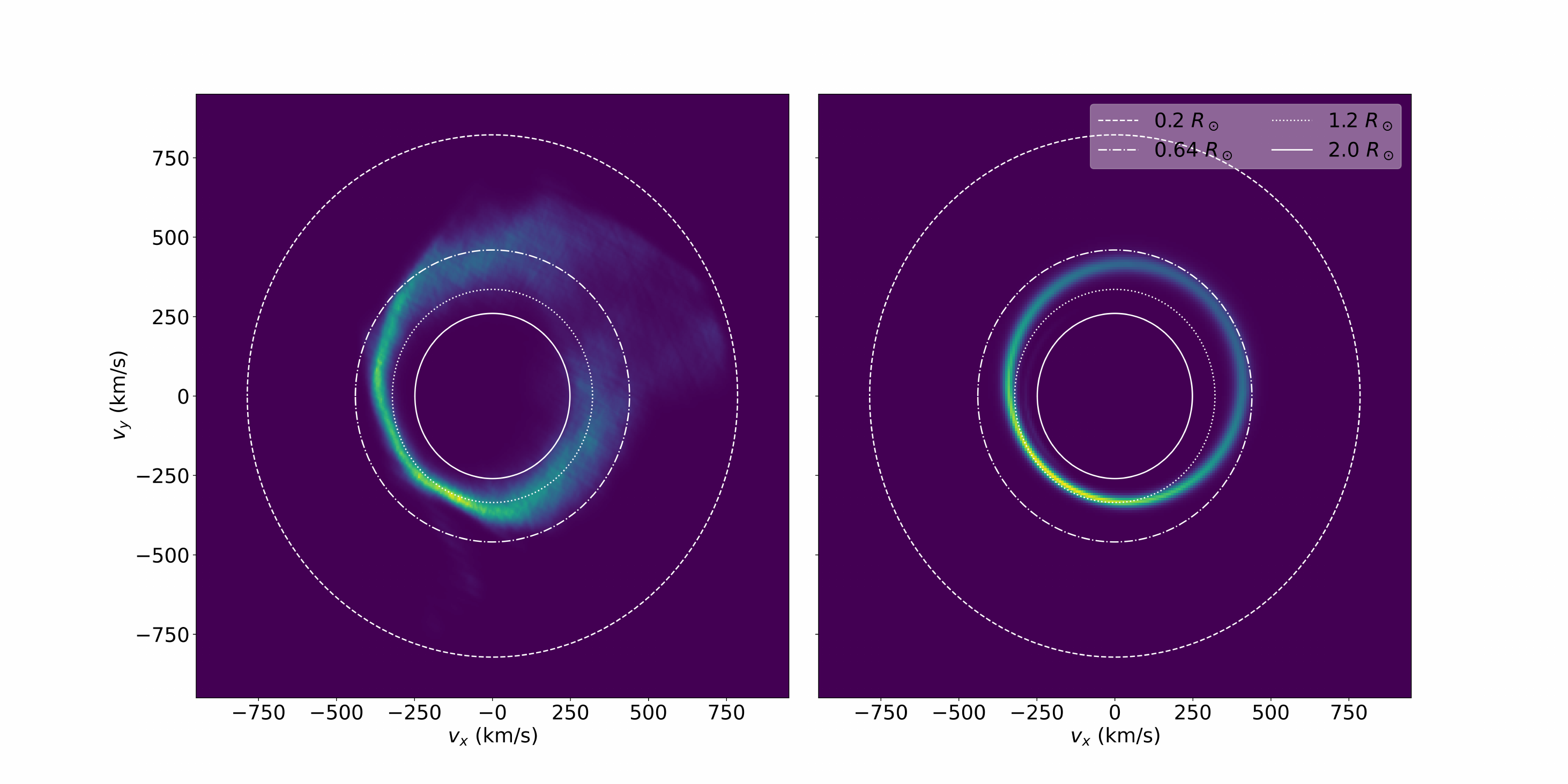}
    \caption{Comparison of tomograms from SDSS J1228+1040 (left), reproduced using data presented in Figure 5 of Manser et al. (2016a) and a simulated gas disc (right). We found a best fit orbit with a semi major axis of $0.920\pm0.005$\,R$_\odot$ and an eccentricity of $0.188\pm 0.004$ from the data for SDSS J1228+1040, used to set the initial orbit of the point mass in the simulated disc. White circles indicate circular Keplerian orbits corresponding to radii of $0.2$\,R$_\odot$, $0.64$\,R$_\odot$, $1.2$\,R$_\odot$, and $2.0$\,R$_\odot$ (outer to inner circles, respectively), inclined by 73 degrees to the observer. High density regions are yellow, while low density regions are light blue.}
    \label{fig:tom_comp}
\end{figure*}

Figure~\ref{fig:tom_comp} compares the observational tomogram of the gas disc around SDSS J1228+1040 to our simulated gas disc. We fit an eccentric orbit to the observational data, finding an eccentricity and semi-major axis of $0.188\pm 0.004$ and $0.879\pm0.005$\,R$_\odot$, respectively. We then simulated a gas disc created from a point mass orbiting with the same semi-major axis and eccentricity. The other parameters of the simulation were identical to the previous simulations. Applying the fitting procedure to our simulated gas disc, we found an eccentricity of $0.143\pm 0.010$, and a semi-major axis of $0.996\pm 0.014$\,R$_\odot$.

\begin{figure}
	\includegraphics[width=\columnwidth]{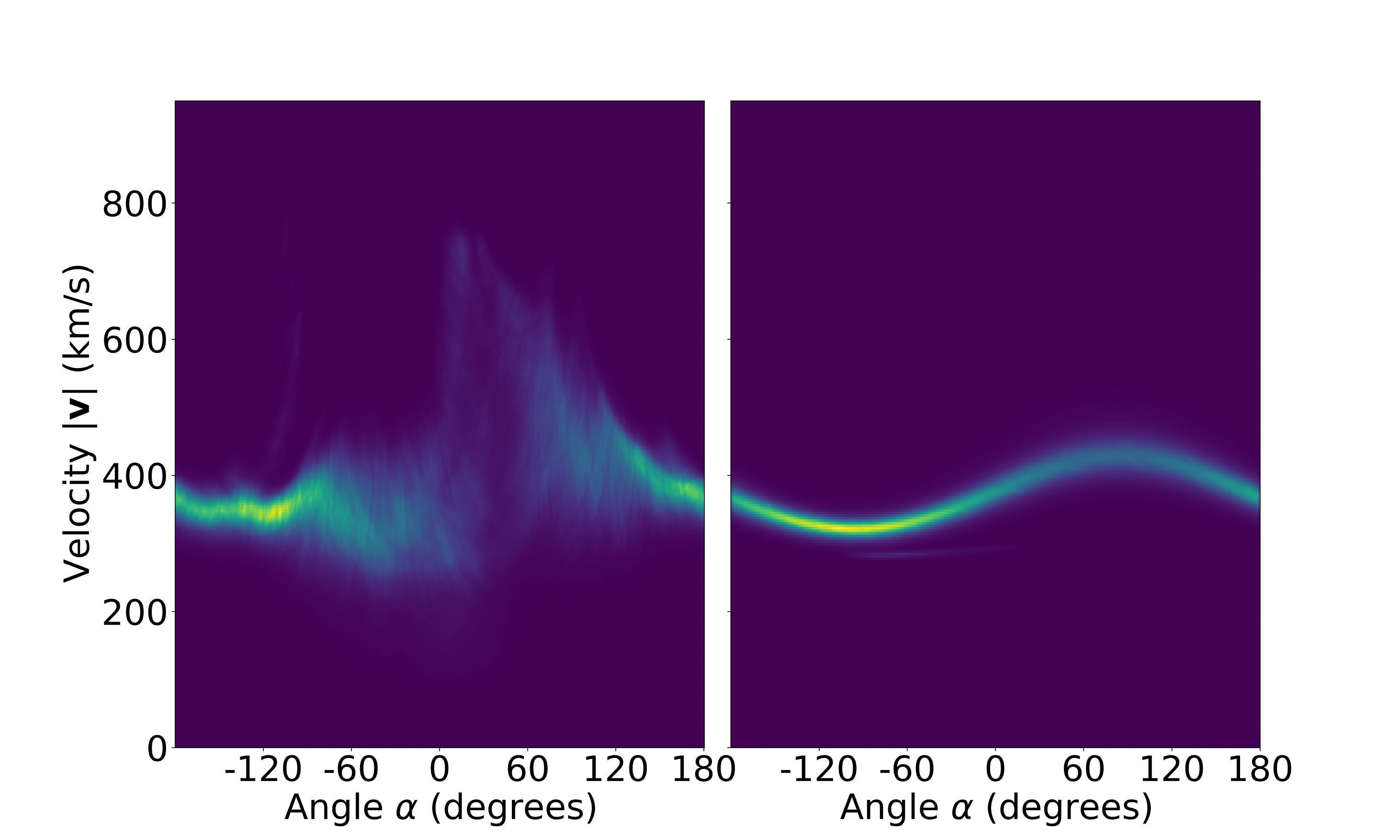}
    \caption{Polar plot of Figure~\ref{fig:tom_comp}. The spread in velocity magnitudes at pericentre in the observational data is $\sim500$\,kms$^{-1}$, whereas in the synthetic data it is under $100$\,kms$^{-1}$. In both plots, we see a concentration of material at apocentre with a spread of $\sim135$ degrees.}
    \label{fig:tom_comp_polar}
\end{figure}

Figure~\ref{fig:tom_comp_polar} shows the same two tomograms plotted in polar coordinates. The concentration of gas at apastron shown in Figure~\ref{fig:tom_ecc_comp} is also evident in both tomograms in Figures~\ref{fig:tom_comp} and \ref{fig:tom_comp_polar}. The main discrepancy between our results and the observations is the larger spread of velocities near pericentre in the observations (see discussion).

For our orbital fits we assumed a 73$^\circ$ inclination between the disc and the line of sight \citep{Manser_2016a}. Fixing either the semi-major axis or the eccentricity to the values inferred from \citet{Manser_2016a} produced fits with reduced $\chi_\nu^2=9.97$ and $\chi_\nu^2=22.07$ respectively, whereas allowing them to vary gave $\chi_\nu^2=6.67$.

\subsection{Resolution study}
\begin{figure}
	\includegraphics[width=\columnwidth]{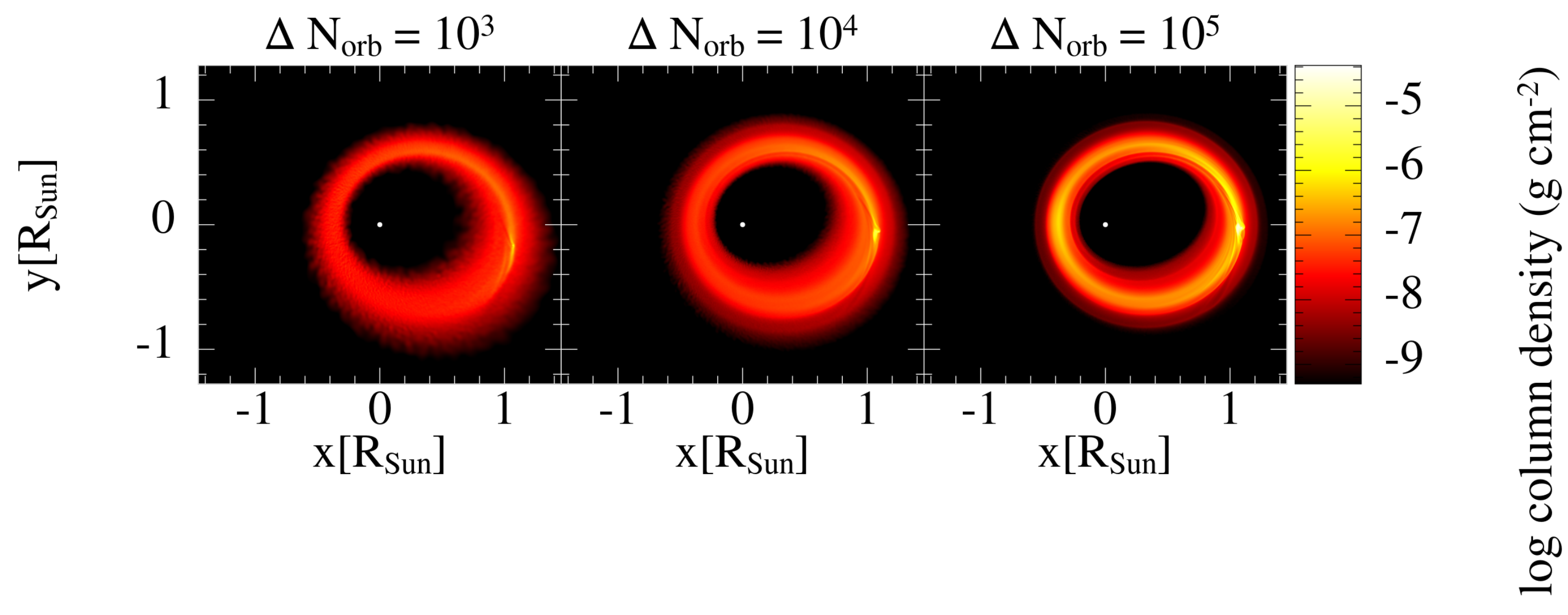}
\vspace{-0.5cm}
    \caption{Column densities of discs created from an orbiting body of eccentricity $e=0.5$ at different particle resolutions (where $\Delta N_{\rm orb}$ is the number of particles injected per orbit). All discs are shown after 100 orbits.}
    \label{fig:res_study}
\end{figure}

Figure~\ref{fig:res_study} shows the column density in three simulations carried out at different numerical resolutions --- changing the mass and hence the number of injected particles by a factor of 10, corresponding to a factor of 2.15 change in smoothing length. The gas disc is more circular at low resolution since the viscosity is higher. 

\section{Discussion}
\label{sec:discussion}
Our results suggest that the presence of eccentric gas discs around at least six white dwarf stars \citep{Wilson_2015, Manser_2016a, Manser_2016b, Dennihy_2018, Dennihy_2020} may be explained by the presence of a body, such as a planet, planetesimal or asteroid, on an eccentric orbit within the disc. Bodies which enter within the tidal radius of the white dwarf (typically $\sim0.005$~au) must come in on highly eccentric orbits, given that the AGB phase of the star consumes planets at up to $\sim1.5$~au \citep{Mustill_2012}. The formation of radially coincident eccentric gas and dust discs \citep{Melis_2010}, is therefore currently best explained by the presence of a partially disrupted body.

The unique detection of a non-disintegrating planetesimal orbiting SDSS J1228+1040 by \citetalias{Manser_2019}, further suggests that we are looking at discs created from partially disrupted bodies. \citetalias{Manser_2019} note that, for this planetesimal to survive to avoid tidal disruption, it would require a density of at least 39 g cm$^{-3}$ without accounting for internal strength, and 7.7 g cm$^{-3}$ while accounting for internal strength. Since iron has a density of 8 g cm$^{-3}$ this suggests a composition similar to the core of a terrestrial or gas giant planet, thus being the remnant of its partial disruption \citep{Shu-lin_2010, Ehrenreich_2015}.

We find that some circularisation occurs for all of the simulated gas discs, exhibited by their decreased eccentricity compared to that of the orbiting body. The short timescale of our simulations does not allow us to explore the long term results of this circularisation. Over longer timescales (years) these discs may reach an equilibrium eccentricity lower than that of the orbital body, where the circularisation of the disc is mediated by the gravitational pull of the orbiting body. The discrepancy between the semi-major axis and eccentricity of the planetesimal detected by \citetalias{Manser_2019}, and the semi-major axis and eccentricity fitted by our model to the gas disc around the same white dwarf, may be the result of this difference in timescales.

The observational data from \cite{Manser_2016a} also exhibits a 500 km s$^{-1}$ spread in velocity at periastron, which is not replicated in any of the simulated discs. Given that the data used by \cite{Manser_2016a} was taken over 12 years, this velocity spread is a feature exhibited on the precession timescale of the disc. Our simulations only explore the evolution of the disc over 100 orbits, so effects present only on the precession timescale, such as velocity spreading at periastron, will not be visible in the simulated discs.


In our simulations, we assume gas is produced directly by the orbiting body. How gas is produced in these discs remains an open question, although several answers have been proposed. \cite{Rafikov_2011} and \cite{Metzger_2012} proposed sublimation of solid debris once it passes within the sublimation radius of the white dwarf. The presence of a planetesimal may increase the sublimation rate of material as dust is perturbed outside of the disc. We assumed $\dot{M} \propto r^{-2}$, consistent with sublimation of volatiles being determined by the light flux from the white dwarf. However, we also assumed gas is emitted isotropically, which is not generally true \citep{Veras_2015b}. We also assumed a gas generation rate equal to the accretion rate of metals measured for SDSS J1228+1040, but the gas injection rate given in Equation \ref{eq:mass inject} is an underestimate of the sublimation rate determined by \citetalias{Manser_2019}. \cite{Kenyon_2017a} and \cite{Kenyon_2017b} proposed that gas is generated through vaporisation of solid debris in collisional cascades. A large solid body, such as a planetesimal, may result in more concentrated vaporisation. A third possibility is the collision of tidally disrupted fragments with preexisting gas or dust \citep{Farihi_2018,Swan_2020, Malamud_2021}. The release rate of gas in these scenarios is uncertain --- see Sections 5.3 and 5.4.1 of \citealt{Malamud_2021} for discussion.


If a non-disrupted planetesimal is necessary for creation of a gas disc around a white dwarf, it may explain the rarity of gas discs among polluted white dwarfs. Gas disc circularisation \citep{Lynden-Bell_1974,Rafikov_2011} is needed for transportation and accretion of metal pollution onto white dwarfs, since the transport of solid material is inefficient \citep{Nixon_2020}. It is therefore striking that between 4 and 30\% of white dwarfs which host debris discs also show emission from gas \citep{Dennihy_2020, GentileFusillo_2020, Manser_2020, Melis_2020}. While there are 21 observed discs with gaseous components orbiting white dwarfs, the number of inferred planetesimals --- either disintegrating or non-disintegrating --- is lower \citep{Vanderburg_2015, Manser_2019, Vanderbosch_2020, Guidry_2020}. We predict stars that host gaseous debris discs should also host planetesimals. Short-term variability studies of the known gaseous discs in emission may thus lead to new detections of such planetesimals.

\vspace{-0.4cm}
\section{Conclusions}
\label{sec:conclusion}
We have demonstrated in this letter that a sublimating or partially disrupting body on an eccentric orbit around a white dwarf forms an eccentric gas disc, that remains eccentric for at least 100 orbits. The disc eccentricity remains within 0.1 of that of the orbiting body.

We found that the eccentricity of the gas disc around the white dwarf SDSS J1228+1040 is $0.188\pm 0.004$, by fitting an orbit to a tomogram of the disc. Simulations performed with these parameters reproduce the observed azimuthal asymmetry and eccentricity.


\vspace{-0.3cm}
\section*{Acknowledgements}
Monash-Warwick Alliance seed funding initiated this project. We thank Rosemary Mardling for useful discussions. DP acknowledges Australian Research Council grant DP180104235. We used the GADI supercomputer, part of the National Computing Infrastructure and also OzStar at Swinburne University. CJM acknowledges an Imperial College Research Fellowship. DV acknowledges an STFC Ernest Rutherford Fellowship (grant ST/P003850/1). RN acknowledges a Stephen Hawking Fellowship from UKRI/EPSRC (EP/T017287/1). We thank the referee for useful review comments.

\vspace{-0.3cm}
\section*{Data Availability}

{\sc phantom} is publicly available and the simulation data is available on request. Data used to produce the Doppler map of SDSS J1228+1040 was reproduced from \cite{Manser_2016a} and will be shared on reasonable request to the corresponding author of \cite{Manser_2016a}.



\bibliographystyle{mnras}
\bibliography{references} 





\label{lastpage}
\end{document}